\documentclass[journal]{IEEEtran}

\normalsize

\usepackage{cite}
\usepackage{graphicx}
\usepackage{subfigure}
\usepackage{url}
\usepackage[centertags]{amsmath}
\usepackage{amsfonts}
\usepackage{amssymb}
\usepackage{dsfont}
\newcommand{\beq}{\begin{equation}}
\newcommand{\eeq}{\end{equation}}
\newcommand{\beqn}{\begin{eqnarray}}
\newcommand{\eeqn}{\end{eqnarray}}

\begin{document}

\title{Generalized Area Spectral Efficiency: An Effective Performance Metric for Green Wireless Communications}
\author{Lei~Zhang,~\IEEEmembership{Student Member,~IEEE,}
        Hong-Chuan~Yang,~\IEEEmembership{Senior Member,~IEEE,}\\
        and~Mazen~O.~Hasna,~\IEEEmembership{Senior Member,~IEEE}
\thanks{Manuscript received February 17, 2013; revised July 14, 2013 and October 27, 2013; accepted December 9, 2013.
This work was supported by startup funds from the University of Victoria, a Discovery Grant from NSERC and a grant from Qatar Telecom (Qtel).}%
\thanks{Lei Zhang and Hong-Chuan Yang are with the Department of Electrical and Computer Engineering, University of Victoria, Victoria, BC, Canada. Email: \textit{leizhang,~hy@uvic.ca}.}%
\thanks{Mazen O. Hasna is with Department of Electrical Engineering, Qatar University, Doha, Qatar. Email: \textit{hasna@qu.edu.qa.}}
}

\markboth{IEEE Transactions on Communications}%
{Submitted paper}

\maketitle

\begin{abstract}
Area spectral efficiency (ASE) was introduced as a metric to quantify the spectral utilization efficiency of cellular systems.
Unlike other performance metrics, ASE takes into account the spatial property of cellular systems.
In this paper, we generalize the concept of ASE to study arbitrary wireless transmissions.
Specifically, we introduce the notion of affected area to characterize the spatial property of arbitrary wireless transmissions.
Based on the definition of affected area, we define the performance metric, generalized area spectral efficiency (GASE), to quantify the spatial spectral utilization efficiency as well as the greenness of wireless transmissions.
After illustrating its evaluation for point-to-point transmission, we analyze the GASE performance of several different transmission scenarios, including dual-hop relay transmission, three-node cooperative relay transmission and underlay cognitive radio transmission.
We derive closed-form expressions for the GASE metric of each transmission scenario under Rayleigh fading environment whenever possible.
Through mathematical analysis and numerical examples, we show that the GASE metric provides a new perspective on the design and optimization of wireless transmissions, especially on the transmitting power selection.
We also show that introducing relay nodes can greatly improve the spatial utilization efficiency of wireless systems.
We illustrate that the GASE metric can help optimize the deployment of underlay cognitive radio systems.
\end{abstract}

\section{Introduction}
Wireless communication systems are carrying an increasing amount of multimedia traffics and, therefore, will have growing ecological impact on our society.
With the current rate of data traffic increase, the energy consumption in wireless networks will grow by approximately 20$\%$ per year \cite{Forster2009}.
To slow down the resulting increase of carbon dioxide emission and achieve more sustainable development, future wireless communication systems need to operate in a more energy efficient fashion.
Various transmission schemes and implementation structures are being developed to support high-data-rate transmissions over limited radio spectrum with the minimum amount of power consumption, which is essentially the ultimate goal of green wireless communications.
To evaluate and compare the ``greenness'' of various transmission schemes, we need effective performance metrics that can characterize the utilization efficiency of both radio spectrum and energy resource \cite{ChenWCSP2010}.

Wireless transmissions generate electromagnetic pollution to the surrounding environment over its operating spectrum band. The size of the polluted area depends on the transmitting power, the radiation pattern of transmit antenna, propagation environment, etc.
In general, if a particular frequency band is heavily ``polluted'', i.e., a significant level of transmitted signal power is observed, over a certain area, simultaneous transmission over the same frequency band in the area may suffer high interference level.
Neighboring transceiver can function properly only when enjoying high signal to interference ratio (SIR) or equipped with effective interference mitigation capability.
As such, an alternative design goal for future green wireless systems is to \emph{achieve high-data-rate transmissions with minimum electromagnetic pollution in both spectral and spatial dimensions}.
To compare the effectiveness of different wireless transmission schemes in approaching such goals, we need a performance metric that can take into account this spatial effect of radio transmissions in the evaluation of transmission efficiency.

Most conventional performance metrics for wireless transmissions focus on the quantification of either spectrum utilization efficiency or link reliability.
In particular, ergodic capacity and average spectrum efficiency evaluate the spectral efficiency of wireless links, where the former serves as the upper bound of achievable average spectrum efficiency \cite{Goldsmith}.
The link reliability is usually quantified in terms of outage probability, average error rate, and average packet loss rate \cite{Verdu2002, Wu2011}.
On another front, various energy efficiency metrics have been developed and investigated, mainly at the component and equipment level \cite{ChenWCSP2010}. In the system/network level, ETSI defined energy efficiency metric as the ratio of coverage area over the power consumption at the base station \cite{etsi09}.  Recently, area power consumption (W/Km$^2$) is introduced and used to optimize base station deployment strategies for cellular network in \cite{Richter09}.
Note that most energy efficiency metrics can not be readily related to the link spectral efficiency or reliability, as they often conflict with each other.
For example, a general tradeoff framework between energy efficiency and spectral efficiency for OFDMA systems was built to characterize their relationship in \cite{Xiong2011}.
Bit per Joule (bit/J), defined as the ratio of achievable rate over the power consumption, is widely used performance metric to quantify the energy utilization efficiency of emerging wireless systems \cite{chen_commag11}.
Recently, this performance metric is applied to the analysis of CoMP cellular systems as well as heterogeneous network \cite{Fehske10, Lorincz12}.
Still, Bit per Joule metric does not take into account the spatial effect of wireless transmissions and is mainly applicable to cellular networks.

Area spectral efficiency (ASE) performance metric was introduced in \cite{Alouini1999} to quantify the spectrum utilization efficiency of cellular systems.
ASE is defined as the maximum data rate per unit bandwidth at a user randomly located in cell coverage area over which the same spectrum is used, with unit being $\text{bps}/(\text{Hz}\cdot \text{m}^2)$.
As co-channel cells in cellular system are separated by a minimum reuse distance of $D$, the same spectrum will be used only once over an area of the size of $\pi D^2/4$.
Note that the area used in ASE definition is based on the co-channel interference requirement of cellular system, not related to the properties of target radio transmissions.
Recently, the ASE performance metric was applied to characterize the performance of two-tier heterogeneous cellular networks \cite{Chandrasekhar2009, Kim2010, Bendlin2011, Ben2012}.
The authors in \cite{Yamamoto2008} studied the ASE performance of cellular systems with cooperative relaying transmission.
Meanwhile, ASE was also analyzed jointly with area power consumption (APC) from a green communication perspective \cite{Zhao2011, Kyuho2011, Arshad2012}.
In these works, ASE was calculated by using either the cell area covered by the macro base station (BS) or \emph{the intensity of the femtocells per unit area according to a certain distribution}.
The effect of smaller ``footprint'' of microcell/femtocell due to lower transmission power were not considered.
After all, the application of ASE metric was still limited to infrastructure based cellular systems.

In this paper, we generalize the ASE metric and develop a new performance metric to evaluate the spectrum efficiency as well as transmission power efficiency of arbitrary wireless transmissions.
The new performance metric, termed as generalized area spectral efficiency (GASE), is defined as the ratio of overall effective ergodic capacity of the transmission link under consideration over the affected area of the transmission.
The \emph{affected area} is defined as the area where a significant amount of transmission power is observed and parallel transmissions over the same frequency will suffer high interference level.
The affected area characterizes the negative effect of radio transmission in terms of electromagnetic pollution while transmitting information to target receivers.
Note that any wireless transmission will generate interference to neighboring transceivers if they operate over the same frequency band.
We use affected area to quantify such spatial effect of wireless transmissions.
The affected area is directly related to the properties of target transmission and makes the GASE metric applicable to arbitrary transmissions.
Note that the size of the affected area depends on various factors, including transmission power, propagation environment, as well as antenna radiation patterns.
In particular, the average radius of the affected area will be proportional to the transmission power.
Therefore, GASE also characterizes the transmission power utilization efficiency in achieving per unit bandwidth throughput and, as such, serves as a suitable quantitative metric for measuring the greenness of wireless communication systems.

\subsection{Contribution, Organization and Notation}
In this paper, we present a comprehensive study on the GASE metric for various transmission scenarios of practical interest.
We first formally introduce the definition of GASE by illustrating its evaluation for conventional point-to-point transmission.
Then we extend the analysis to three transmission scenarios, namely dual-hop relay transmission \cite{meulen71, cover79, nabar04, Laneman}, three-node cooperative relay transmission \cite{Ge2011, Huynh2009, Song2006, Li2012, Adachi2012, Brown2011, Gao2010}, and underlay cognitive radio transmission \cite{Mitola2000, FCC2002, Weiss2004, Goldsmith2009, Haykin2005, Gastpar2007, Ghasemi2007, Musavian2009, Zhang2009, Wang2009}. Cognitive radio has received significant attention lately as it can help greatly improve the spectrum utilization of licensed frequency bandwidth \cite{Mitola2000, FCC2002, Weiss2004}.
Typically, there are three main cognitive radio paradigms \cite{Goldsmith2009}: interweave, overlay and underlay. 
With the underlay paradigm, the secondary cognitive users can access the frequency bandwidth of the primary radio only if the resultant interference power level at the primary receiver is below a given threshold \cite{Haykin2005, Gastpar2007}.
Recently, intensive research has been carried out to quantify the capacity gains of underlay cognitive radio transmission \cite{Ghasemi2007, Musavian2009, Zhang2009, Wang2009}.
These research focused on the benefit of spectrum sharing by imposing a interference constraint on the primary receiver.
However, the spatial property of parallel radio transmissions are overlooked in these analysis.
In particular, the area affected by simultaneous transmission should also be considered when evaluating overall system spectrum utilization efficiency, especially in dense frequency reuse scenario.
For each communication scenario, the generic formula of GASE are presented with specific closed-form expressions for Rayleigh fading environment derived whenever feasible.
Selected numerical examples are presented and discussed to illustrate the mathematical formulation and demonstrate the new insights that GASE metric brings into wireless system design.

The key contributions of this paper are summarized as follows:
\begin{itemize}

\item
We introduce the concept of affected area to arbitrary wireless communications, based on which we present a new performance metric GASE, which can evaluate the spectrum utilization efficiency as well as power utilization efficiency of arbitrary wireless transmissions.
We observe that while the conventional spectral efficiency performance metric is a monotonically increasing function of the transmission power, the GASE metric is no longer a monotonic function of transmission power.
In fact, optimal transmission power value exists in terms of maximizing the GASE of wireless transmissions.
We also notice that larger transmission power leads to higher spectral efficiency but not necessarily higher GASE. As such, the GASE metric provides a new perspective on the design and optimization of wireless systems.

\item
We study the GASE performance of dual-hop relay transmission and three-node cooperative relay transmission scenarios with either AF or DF relaying modes.
Through the analytical results and selected numerical examples, we demonstrate that relay transmission can achieve higher maximum GASE with smaller transmission power than point-to-point transmission.
In addition, if the transmission power is adjusted properly, relay transmission enjoys better overall GASE performance than that of point-to-point link.
Therefore, relay transmission can effectively improve the power efficiency of wireless transmission.

\item
We derive the accurate analytical expressions of GASE for underlay cognitive radio transmission, which includes that for point-to-point transmission and $X$ channel transmission as limiting special cases.
Through a comparative study with the overall spectral efficiency metric, we develop new design guidelines for underlay cognitive radio implementation.
We show that GASE performance metric provides a new perspective on the design of underlay cognitive radio transmission.
Specifically, the overall spectral efficiency with underlay cognitive transmission can not be worse than that of the point-to-point primary transmission only case.
However, the GASE performance can be much worse than that of point-to-point transmission if the power of secondary transmitter is not properly chosen.
The transmission power of secondary cognitive transmission should be carefully selected to benefit overall GASE performance of underlay cognitive system, especially when the interfering link distance is considerably greater than the desired transmission link distance.

\end{itemize}

The remainder of this paper is organized as follows:
Section \uppercase \expandafter {\romannumeral 2} introduces the definition of GASE by considering conventional point-to-point transmission.
Section \uppercase \expandafter {\romannumeral 3} generalizes the analysis of GASE to dual-hop relay transmission while in
Section \uppercase \expandafter {\romannumeral 4}, we consider GASE of three-node cooperative relay transmission.
Section \uppercase \expandafter {\romannumeral 5} analyzes GASE of underlay cognitive radio transmission and shows its asymptotic characteristics.
Finally, section \uppercase \expandafter {\romannumeral 6} concludes the paper.

Throughout this paper, $F_X(x)$ denotes the cumulative distribution function (cdf) of random variable (RV) $X$.
$f_X(x)$ denotes the probability density function (pdf) of $X$.
$\mathcal{E}(\lambda)$ denotes the exponential distribution with mean $1/\lambda$.
$\mathbb{P} \left\{ \cdot \right\}$ denotes the probability of a random event.

\section{Generalized Area Spectral Efficiency}
In this section, we formally introduce the definition of GASE metric and illustrate its evaluation for conventional point-to-point transmission.
Consider a point-to-point wireless link between generic source $\text{S}$ and destination $\text{D}$.
The source $\text{S}$ is transmitting with power $P_t$ using an omni-directional antenna.
For analytical tractability and presentation clarity, we assume that the transmitted signal experiences both path loss and multipath fading effects, ignoring the shadowing effect.
Specifically, the received signal power $P_{r}$ at distance $d$ from the transmitter is given by
$
\label{propagation_model2}
P_{r} = P_{t}\cdot Z / (d/d_{\text{ref}})^{a},
$
where $a$ is the path loss exponent, and $Z$ is an independent RV that models the  multipath fading effect, $d_{\text{ref}}$ is the reference distance.
Without generality, we set $d_{\text{ref}} = 1$m.
We also assume that the fading channel is slowly varying and, as such, the transmitter can adapt its transmission rate with the channel condition for reliable transmission.

GASE is defined as the ratio of the ergodic capacity of the link, denoted by $\overline{\text{C}}$, over the size of the affected area of the transmission, denoted by $\text{A}$.
Mathematically, if we denote GASE by $\eta$, we have
$
\eta = \overline{\text{C}} / \text{A}.
$
The affected area refers to the area where a significant amount of transmission power is observed, i.e., the received signal power $P_r$ is greater than a certain threshold value $P_{\text{min}}$, which will lead to significant amount of interference to neighboring transceivers.
{ The value of $P_{\text{min}}$ should be selected based on the interference sensitivity of neighboring wireless transmissions.
In particular, if the neighbouring transmission is insensitive to interference, e.g., spread spectrum or ultra wideband (UWB) systems, $P_{\text{min}}$ may be set to a large value. Otherwise, it should be set to the same order of magnitude to background noise. The same $P_{\text{min}}$ value should be used to compare different design for the target transmission. }
For point-to-point link, the probability that an incremental area of distance $r$ from the transmitter is affected is equal to the probability that the received signal power, $P_r(r)$, is greater than $P_{\text{min}}$, i.e., $\mathbb{P} \big\{ P_t \cdot Z / r^a \ge P_{\text{min}} \big\}$. 
It follows that the affected area of point-to-point transmission can be determined as
\beqn
\label{Aaffgen}\nonumber
\text{A} &=& \int_0^{2\pi} \int_{0}^{\infty} \int_{ P_{\text{min}} \cdot r^a / P_t }^{\infty} f_Z (z) \, \mathrm{d} z \, r \mathrm{d} r \mathrm{d} \theta \\
&=&2\pi \int_{0}^{\infty} \bigl(1-F_Z \bigl(P_{\text{min}} \cdot r^a / P_t\bigr) \bigr) \,  r \mathrm{d} r.
\eeqn
Meanwhile, the instantaneous link capacity between source $\text{S}$ and destination $\text{D}$ is given by
$
\text{C} = \log_2 \left( 1 + \frac{P_t }{N d^a } \cdot z \right),
$
where $z$ denotes a particular realization of fading power gain $Z$ and $N$ is the noise power.
As such, the ergodic capacity can be calculated by averaging the instantaneous link capacity over the distribution of $Z$.
Mathematically speaking, we have
\beq
\label{capacityforsingleuserdef}
\overline{\text{C}} = \int_0^\infty \log_2 \bigg( 1 + \frac{P_t }{N d^a } \cdot z \bigg) \, \mathrm{d} F_Z(z),
\eeq
where $F_{Z}(\cdot)$ is the cdf of $Z$. Note that we ignore the effect of external interference from neighbouring transmission in non affected area in the capacity calculation. We assume that if $P_{\text{min}}$ is chosen properly, due to the channel reciprocity property, the neighbouring transmission from non affected area will not generate significant interference to the target transmission. The effect of such external interference on the ergodic capacity as well as GASE will be addressed in our future work.

Therefore, the GASE for point-to-point link can be calculated as
\begin{equation}
{\eta} =\frac{\int_0^\infty \log_2 \big( 1 + \frac{P_t }{N d^a } \cdot z \big)  \, \mathrm{d} F_{Z}(z)}{2\pi \int_{0}^{\infty} \bigl(1-F_Z \bigl(P_{\text{min}} \cdot r^a / P_t\bigr) \bigr) \,  r \mathrm{d} r}.
\label{GASE_avg_defforsingleuser}
\end{equation}

Under Rayleigh fading environment, $Z$ is an exponential RV with unit mean, i.e., $Z \thicksim \mathcal{E}(1)$. The affected area of point-to-point transmission specializes, with the help of \cite[Eq.  3.326.2]{Tables}, to
\beq
\label{AaffgenSingleUser}
\text{A} = \frac{2\pi}{a}\Gamma\left(\frac{2}{a}\right) \left(\frac{P_{t}}{ P_{\text{min}}}\right)^{2/a},
\eeq
where $\Gamma(\cdot)$ denotes the Gamma function.
As we can see from \eqref{AaffgenSingleUser},  the affected area for the point-to-point link over Rayleigh fading is proportional to  $P_t^{2/a}$, where $a$ is the path loss exponent. Meanwhile, the ergodic capacity of the point-to-point link is given by
\beq
\label{capacityforsingleuser}
\overline{\text{C}}
=\frac{1}{\ln 2} \operatorname{E}_1\left(\frac{d^a N}{P_t}\right) \exp\left(\frac{d^a N}{P_t}\right),
\eeq
where $\operatorname{E}_1(x) = \int_x^{\infty} \frac{e^{-t}}{t} \mathrm{d} t$ is the exponential integral function \cite{Tables}.
Finally, we can obtain the closed-form expression of GASE for point-to-point transmission over Rayleigh fading as
\begin{equation}
\eta = \frac{\frac{1}{\ln 2} \operatorname{E}_1\left(\frac{d^a N}{P_t}\right) \exp\left(\frac{d^a N}{P_t}\right)}{\frac{2\pi}{a}\Gamma\left(\frac{2}{a}\right) \left(\frac{P_{t}}{ P_{\text{min}}}\right)^{2/a}}.
\label{GASEforsingleuserray}
\end{equation}
It worths noting that, by including the factor $P_t^{2/a}$ in the denominator, the GASE performance metric also quantifies the energy utilization efficiency of wireless transmissions in achieving certain ergodic capacity while taking into account radio propagation effects. The conventional bit per Joule metric, specialized to $C/P_t$ for point-to-point transmission, is roughly equivalent to the special case of GASE metric with $a=2$.

\begin{figure}[h]
\begin{center}
\includegraphics[width = 3 in] {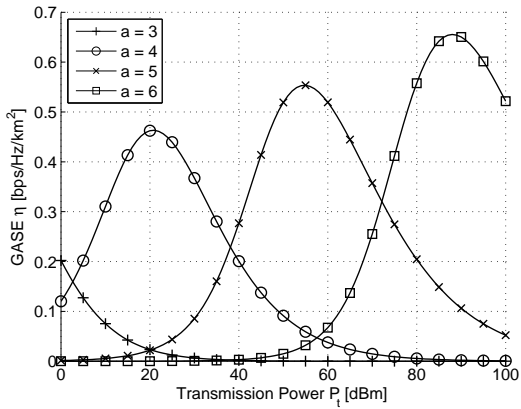}\\
\end{center}
\caption{The effect of transmission power $P_t$ on $\eta$. $N = -100$ dBm, $P_{\text{min}} = -90$ dBm, $d = 1000$ m.}
\label{GASE_SingleUser}
\end{figure}

In Fig. \ref{GASE_SingleUser}, we plot the GASE of point-to-point link under Rayleigh fading environment as function of the transmission power $P_t$ for different path loss exponent $a$.
It is interesting to see that, unlike conventional spectral efficiency metric, the GASE $\eta$ is not a monotonic function of $P_t$ in general.
Only when $a$ is very small will GASE $\eta$ be a monotonically decreasing function of the transmission power $P_t$.
For medium to large $a$, GASE is proportional to $P_t$ when $P_t$ is relatively small, which implies that  the ergodic capacity increases faster than the affected area in this region.
When $P_t$ becomes large, the affected area increases faster, which leads to a decreasing GASE.
As such, an optimal $P_t$ value exists in terms of maximizing the GASE of point-to-point transmission.
These behaviors of $\eta$ can be mathematically verified with the following limiting results based on \eqref{GASEforsingleuserray},
\begin{equation}
\lim_{P_t\rightarrow 0^+} \eta = \begin{cases}
\infty, & a<2;\\
\log_2 e \cdot \frac{P_{\text{min}}}{\pi N d^2}, & a = 2;\\
0, & a>2,
\end{cases}
\label{GASEllimitsforsingleuserray}
\end{equation}
and
\begin{equation}
\lim_{P_t\rightarrow \infty} \eta = 0.
\label{GASEulimitsforsingleuserray}
\end{equation}
It is straightforward although tedious to verify that $\eta$ is a concave function of $P_t$. As such, there exist optimal values for $P_t$ that maximize the GASE for point-to-point link for $a > 2$ cases, which can be analytically obtained by solving $\frac{d}{d P_t} \eta =  0$ for $P_t$.
After substituting \eqref{GASEforsingleuserray} into it and some manipulations, we arrive at the following equation that the optimal $P_t^*$ satisfies
\beq
\bigg( \frac{d^a N}{P_t^*} + \frac{2}{a} \bigg) \operatorname{E}_1\left(\frac{d^a N}{P_t^*}\right) \exp\left(\frac{d^a N}{P_t^*}\right) = 1.
\eeq
Various numerical methods can be used to solve this integral equation for $P_t^*$.
Note that the optimal transmitting power value is proportional to the product $d^a N$, but independent of the minimum power threshold $P_{\text{min}}$.
Essentially, for a certain propagation environment and source-destination distance, $P_t^*$ leads to the largest ergodic capacity per unit affected area.
The transmission power is therefore optimally utilized with consideration of the spatial effect of radio transmission.

Fig. \ref{GASE_SingleUser} also shows that the larger the path loss exponent $a$, the larger the maximum achievable GASE.
Meanwhile, we need to use higher transmission power to achieve this maximum GASE.
This observation indicates that when the path loss effect is significant, it is beneficial to use high transmission power as the affected area is not growing quickly.
On the other hand, when there is severe shadowing effect between the $\text{S}$-$\text{D}$ link, the GASE performance will suffer from the excessive increase of the transmission power $P_t$.
In such scenario, relay transmission is the ideal solution to increase link throughput without significantly increasing the spatial footprint.
In the next section, we examine the GASE performance of relay transmission.

\section{GASE of Dual-Hop Relay Transmission}
\begin{figure}[h]
\begin{center}
    \includegraphics[width = 2 in] {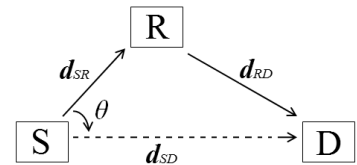}\\
\end{center}
\caption{Dual hop relay transmission. $\text{S}$, $\text{D}$ and $\text{R}$ represent the information source, destination and relay node, respectively.}
\label{RelayCase}
\end{figure}
In this section, we consider the scenario where $\text{S}$ transmits data to $\text{D}$  with the help of the intermediate relay node $\text{R}$, as illustrated in Fig. \ref{RelayCase}. Specifically, relay $\text{R}$ carries out either decode-and-forward (DF) or amplify-and-forward (AF) operation in a half-duplex mode.
We assume the relay transmission occurs in two successive time slots of equal duration $T$.
The distance from source to relay and relay to destination are denoted by $d_{\textit{SR}}$ and $d_{\textit{RD}}$, respectively. We assume that direct $\text{S}$-$\text{D}$ transmission is not possible, due to for example deep shadowing. The GASE performance of cooperative three-node network will be considered in the next section.

The transmission power of the source and relay node are denoted by $P_{S}$ and $P_{R}$, respectively.
The ergodic capacity of the relay transmission can be calculated, noting the half-duplex constraint, as
\beq
\label{CR}
\overline{\text{C}}_{R} = \frac12 \int_{0}^{\infty} \log_2 (1 + \gamma) \cdot f_{\Gamma_{\text{eq}}} (\gamma) \, \mathrm{d} \gamma,
\eeq
where $f_{\Gamma_{\text{eq}}}(\cdot)$ denotes the pdf of the equivalent end-to-end SNR $\Gamma_{\text{eq}}$.
For DF relaying protocol, $\Gamma_{\text{eq}}^{\textit{DF}}$ is equal to $\min \left\{\Gamma_{\textit{SR}},  \Gamma_{\textit{RD}} \right\}$, whereas for AF protocol, ${\Gamma_{\text{eq}}^{\textit{AF}}}$ is given by
$
\Gamma_{\text{eq}}^{\textit{AF}} = \frac{\Gamma_{\textit{SR}} \cdot \Gamma_{\textit{RD}}}{\Gamma_{\textit{SR}} + \Gamma_{\textit{RD}} + 1}
$ \cite{Laneman},
where $\Gamma_{\textit{SR}}$ and $\Gamma_{\textit{RD}}$ are the instantaneous received SNR of S-R hop and R-D hop, respectively.
The affected area for the first and second relay transmission step, denoted as $\text{A}_{\textit{SR}}$ and $\text{A}_{\textit{RD}}$, can be calculated using \eqref{AaffgenSingleUser} with correspondent transmission power $P_{S}$ and $P_{R}$, respectively. Note that $\text{A}_{\textit{SR}}$ and $\text{A}_{\textit{RD}}$ will not be affected at the same time as $\text{S}$ and $\text{R}$ transmit alternatively.
Therefore, the overall GASE for dual-hop relay transmission is calculated by averaging GASE of source and relay transmission steps, while noting that each step finishes half of data transmission, as
\beq
\label{eta_relaygen}
{\eta}_R^{ } = \frac12 \bigg\{ \frac{\overline{\text{C}}_{R}}{\text{A}_{\textit{SR}}} + \frac{\overline{\text{C}}_{R}}{\text{A}_{\textit{RD}}}\bigg\}.
\eeq

Under Rayleigh fading environment, the received SNR $\Gamma_{ij}$ ($i \in\{S, R\}, j \in \{R, D\}$ and i $\neq j$) can be expressed as
$\Gamma_{ij} = \bar{\gamma}_{ij} \cdot Z,$
where $\bar{\gamma}_{ij} = \frac{P_i}{d_{ij}^a \cdot N}$ is the average received SNR related to the distance from the transmitter $i$ to receiver $j$, $d_{ij}$.
$Z$ is an exponential RV with unit mean.
It follows that the pdf of $\Gamma_{\text{eq}}^{\textit{DF}}$ can be obtained, noting that $\Gamma_{\text{eq}}^{\textit{DF}}$ is the minimum of two exponential RVs, as
\beq
\label{PDF_Gamma_eq_DF}
f_{\Gamma_{\text{eq}}^{\textit{DF}}} (\gamma) = \alpha_{1} \cdot e^{-\alpha_{1} \gamma},
\eeq
where $\alpha_{1} = \frac{1}{\overline{\gamma}_{\textit{SR}}} + \frac{1}{\overline{\gamma}_{\textit{RD}}}$.
Then the ergodic capacity for DF case can be derived as
\begin{align}
\label{CDF}
\overline{\text{C}}_{\textit{DF}}
&= \frac{1}{2 \ln 2} \operatorname{E}_1 (\alpha_1) \exp (\alpha_1).
\end{align}
Finally, GASE of DF relay transmission for Rayleigh fading scenario is given, after applying \eqref{AaffgenSingleUser} and \eqref{CDF} into \eqref{eta_relaygen}, by
\beq
\label{eta_relaydf}
{\eta}^{\textit{DF}}_{R} = \frac{1}{4\ln2} \left\{ \frac{\operatorname{E}_1 (\alpha_1) \exp (\alpha_1)}{\frac{2\pi}{a}\Gamma\left(\frac{2}{a}\right) \left(\frac{P_{S}}{ P_{\text{min}}}\right)^{2/a}} + \frac{\operatorname{E}_1 (\alpha_1) \exp (\alpha_1)}{\frac{2\pi}{a}\Gamma\left(\frac{2}{a}\right) \left(\frac{P_{R}}{ P_{\text{min}}}\right)^{2/a}}\right\}.
\eeq

For AF relaying, the pdf of the equivalent SNR $\Gamma_{\text{eq}}^{\textit{AF}}$ is approximately obtained as \cite[eq. (18)]{Hasna2002}
\beq
\label{PDF_Gamma_eq_AF}
f_{\Gamma_{\text{eq}}^{\textit{AF}}} (\gamma) = 2\beta_1 \gamma e^{-\alpha_1 \gamma} \bigg\{ \alpha_1 \operatorname{K}_1(2\beta_1 \gamma) + 2\beta_1 \operatorname{K}_0 (2\beta_1 \gamma)\bigg\},
\eeq
where $\beta_1 = \frac{1}{\sqrt{\overline{\gamma}_{\textit{SR}} \cdot \overline{\gamma}_{\textit{RD}}}}$, $\operatorname{K}_0(\cdot)$ and $\operatorname{K}_1(\cdot)$ is the second kind modified Bessel function of the zero-order and first-order, respectively \cite{Tables}.
Substituting  \eqref{PDF_Gamma_eq_AF} into \eqref{CR}, we can calculate ergodic capacity of relay transmission with AF protocol as
\begin{align}
\label{CAF}
\nonumber
\overline{\text{C}}_{\textit{AF}}
= \int_{0}^{\infty} & \log_2  (1 + \gamma) \beta_1 \gamma e^{-\alpha_1 \gamma} \times \\
&\Big\{ \alpha_1 \operatorname{K}_1(2\beta_1 \gamma) + 2\beta_1 \operatorname{K}_0 (2\beta_1 \gamma)\Big\} \mathrm{d}\gamma.
\end{align}
The GASE performance of AF-based relay transmissions can be similarly calculated by applying \eqref{CAF}  in \eqref{eta_relaygen}.

\begin{figure}[h]
\begin{center}
    \includegraphics[width = 3.5 in] {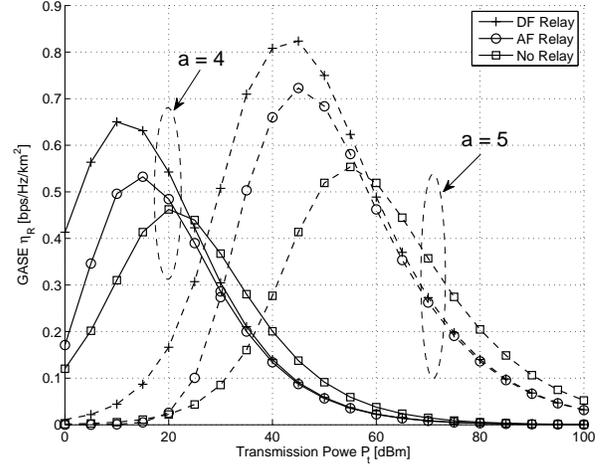}\\
\end{center}
\caption{Comparison of the dual-hop relay transmission and point-to-point transmission. $N = -100 $ dBm, $P_{\text{min}} = -90 $ dBm, $d_{\textit{SD}} = 1000 $ m, $d_{\textit{SR}} = d_{\textit{RD}} = 500$ m, $\theta = 0$.}
\label{GASE_DF_vs_AF_vs_SingleUser}
\end{figure}

In Fig. \ref{GASE_DF_vs_AF_vs_SingleUser}, we compare the GASE of dual-hop relay transmission with point-to-point transmission.
Specifically, the GASE of both DF and AF cases are plotted as function of common transmission power $P_{S}=P_{R} = P_t$ for different values of path loss exponent $a$.
The relay node $\text{R}$ is assumed to be at the center point along the line between $\text{S}$ and $\text{D}$.
As we can see, similar to the point-to-point transmission case, there exist optimal values for transmitting power $P_t$ in terms of maximizing GASE of dual-hop relay transmission.
Based on the analytical results on GASE, the optimal transmitting power for relay transmissions can be obtained by solving the following optimization problem
\begin{align}
\label{opt2D}
& \max_{P_{S}, P_{R} \in {\mathbb R}^+}  \eta_{R}\\\nonumber
s. t.\ \ \   & P_{S} < P_{\text{max}},  P_{R} < P_{\text{max}},
\end{align}
where $P_{\text{max}}$ is the maximum transmission power of the nodes.
For DF relaying under Rayleigh fading environment, after applying \eqref{CDF} and \eqref{AaffgenSingleUser}, the objective function specializes to
\begin{align}
{\eta}_{R}^{\textit{DF}} =& \frac{a}{8\pi \ln 2 \cdot \Gamma\left(\frac{2}{a}\right)} \operatorname{E}_1 (\alpha_1) \exp (\alpha_1)\\\nonumber
& \times \left(\left(\frac{P_{S}}{ P_{\text{min}}}\right)^{-2/a}+ \left(\frac{P_{R}}{ P_{\text{min}}}\right)^{-2/a}\right).
\end{align}
The resulting optimization problem can be easily solved numerically.
We also note from Fig. \ref{GASE_DF_vs_AF_vs_SingleUser} that the maximum GASE of relay transmission are much higher and can be achieved with much smaller transmission power than point-to-point transmission.
From this perspective, introducing a relay node can greatly improve the greenness of wireless transmissions.
On the other hand, if the transmission power are set too large, the GASE of point-to-point transmission becomes slightly larger than that of relay transmission, partly due to the half duplex constraint on relay transmission.
Therefore, dual-hop relay transmission is more energy efficient than point-to-point transmission only when the transmission power is adjusted to proper values.

\section{GASE of Cooperative Relay Transmission}
Previous section investigates the GASE performance of relay transmission and ignores the direct source to destination link. In this section, we utilize GASE metric to quantify the spectrum as well as power utilization efficiency of cooperative transmissions while taking into account the spatial effects of each transmission stage.
We focus on a three-node cooperative relay transmission where relaying is activated only if it will lead to higher instantaneous capacity.
In particular, the source node decides to perform either direct or relay transmission to communicate with the destination node based on the instantaneous link capacity.
Therefore, the instantaneous capacity of such three-node cooperative transmission is given by
\beq
\label{maxCinst}
\mathbf{\text{C}}_{\text{inst}} = \max \left\{ \mathbf{\text{C}}_\text{d}, \mathbf{\text{C}}_\text{r}\right\},
\eeq
where $\mathbf{\text{C}}_\text{d}$ and $\mathbf{\text{C}}_\text{r}$ are the instantaneous capacity of direct transmission and relay transmission, respectively.
$\mathbf{\text{C}}_\text{d}$ is related to the instantaneous received SNR of $\text{S}$-$\text{D}$ link, $\Gamma_{\textit{SD}}$, as
\begin{equation}
\label{Cdirect}
\mathbf{\text{C}}_\text{d} = \log_2 (1 + \Gamma_{\textit{SD}}),
\end{equation}
The instantaneous capacity of relay transmission $\mathbf{\text{C}}_\text{r}$ is given by
\begin{equation}
\label{Crelay}
\mathbf{\text{C}}_\text{r} = \frac12 \log_2 (1 + \Gamma_{\text{eq}}),
\end{equation}
where $\Gamma_{\text{eq}}$ is the equivalent received SNR of the relay channel, and the factor $\frac12$ is due to the half-duplex constraint.
Substituting \eqref{Cdirect} and \eqref{Crelay} into \eqref{maxCinst}, the instantaneous capacity specializes to
\beqn
\label{genCinst}
\nonumber
\mathbf{\text{C}}_{\text{inst}} &=& \frac12 \max \bigg\{ \log_2 (1+\Gamma_{\textit{SD}})^2, \log_2 (1+\Gamma_{\textit{eq}}) \bigg\} \\
&=&\frac12 \log_2 \bigg\{ 1 + \underbrace{\max \Big\{ \Gamma_{\textit{SD}}^2 + 2\Gamma_{\textit{SD}}, \Gamma_{\text{eq}} \Big\} }_{\Gamma_{\text{C}}} \bigg\},
\eeqn
where $\Gamma_{\text{C}}$ is the overall equivalent received SNR of the three-node cooperative transmission.
The ergodic capacity can be derived by averaging the instantaneous capacity over the distribution of $\Gamma_{\text{C}}$, i.e.,
\beq
\label{genavgCapacity}
\overline{\mathbf{\text{C}}} = \int_{0}^{\infty} \frac12 \log_2 \big(1 + \gamma \big) \cdot f_{\Gamma_{\text{C}}} (\gamma) \, \mathrm{d} \gamma,
\eeq
where $f_{\Gamma_{\text{C}}} (\gamma)$ is the pdf of $\Gamma_{\text{C}}$.
Meanwhile, the probability that the system performs direct transmission is equal to the probability that $\mathbf{\text{C}}_\text{d} > \mathbf{\text{C}}_\text{r}$, i.e.,
\beq
\mathcal{P}_{\text{d}} = \mathbb{P} \bigg\{\Gamma_{\textit{SD}}^2 + 2\Gamma_{\textit{SD}} >  \Gamma_{\text{eq}} \bigg\}.
\eeq
Accordingly, the probability that the system performs relay transmission is given by $\mathcal{P}_{\text{r}} = 1 - \mathcal{P}_{\text{d}}$.

The GASE of three-node cooperative transmission, denoted by $\eta_{C}$, can be calculated, while noting that the affected areas of source and relay transmissions are different, as
\begin{equation}
\label{defGASE}
\eta_{C} = \mathcal{P}_{\text{d}} \cdot \frac{\overline{\mathbf{\text{C}}}_\text{d}}{\text{A}_{\textit{SR}}} + \mathcal{P}_{\text{r}} \cdot\frac12 \left( \frac{\overline{\mathbf{\text{C}}}_\text{r}}{\text{A}_{\textit{SR}}}+ \frac{\overline{\mathbf{\text{C}}}_\text{r}}{\text{A}_{\textit{RD}}} \right),
\end{equation}
where $\text{A}_{\textit{SR}}$ and $\text{A}_{\textit{RD}}$ are the affected areas for the source and relay transmission steps, which can be calculated using \eqref{AaffgenSingleUser} with correspondent transmission power; $\overline{\mathbf{\text{C}}}_\text{d}$ and $\overline{\mathbf{\text{C}}}_\text{r}$ are the average ergodic capacity under direct and relay transmission.
In what follows, we will calculate $\overline{\mathbf{\text{C}}}_\text{d}$ and $\overline{\mathbf{\text{C}}}_\text{r}$ for DF and AF relaying protocols under Rayleigh fading environment.

\subsection{DF Relaying Protocol}
Under Rayleigh fading environment, $\Gamma_{\textit{SD}} = \overline{\gamma}_{{\textit{SD}}} \cdot Z$ and $Z \sim \mathcal{E}(1)$, $\mathcal{P}_{\text{d}}$ can be specialized to
\beq
\label{genPrdirect}
\mathcal{P}_{\text{d}} = \frac{1}{\overline{\gamma}_{{\textit{SD}}}} \int_{0}^{\infty} F_{\Gamma_{{\text{eq}}}} (x^2+2x) \exp(-x/\overline{\gamma}_{{\textit{SD}}}) \, \mathrm{d} x.
\eeq
With DF relaying, the pdf of equivalent received SNR over relay link, $\Gamma_{\text{eq}}^{\textit{DF}}$, is given by \eqref{PDF_Gamma_eq_DF}.
Substituting \eqref{PDF_Gamma_eq_DF} into \eqref{genPrdirect} and carrying out integration, we can obtain the probability that the system performs direct transmission with DF relaying protocol, as
\beq
\mathcal{P}_{\text{d}}^{\textit{DF}} = 1 - \frac{1}{\overline{\gamma}_{\textit{SD}}} \mathfrak{D}(\alpha_{1},\alpha_{2}),
\eeq
where $\alpha_{1} = \frac{1}{\overline{\gamma}_{\textit{SR}}} + \frac{1}{\overline{\gamma}_{\textit{RD}}}$, $\alpha_{2} = \frac{2}{\overline{\gamma}_{\textit{SR}}} + \frac{2}{\overline{\gamma}_{\textit{RD}}} + \frac{1}{\overline{\gamma}_{\textit{SD}}}$, and $\mathfrak{D}(\alpha_{1}, \alpha_{2})$ is defined as
\begin{align}
\nonumber
\mathfrak{D}( \alpha_{1}, \alpha_{2}) &\triangleq \int_{0}^{\infty} e^{-\alpha_{1} t^2 - \alpha_{2} t} \mathrm{d} t \\
&= \frac12 \sqrt{\frac{\pi}{\alpha_{1}}} e^{\frac{\alpha_{2}^2}{4 \alpha_{1}}} \operatorname{erfc} (\frac{\alpha_{2}}{2\sqrt{\alpha_{1}}}),
\end{align}
where $\operatorname{erfc}(x) = \frac{2}{\sqrt{\pi}} \int_{x}^{\infty} e^{-t^2} \mathrm{d} t$ is the complementary error function \cite{Tables}.

The equivalent SNR of the three-node cooperative network with DF protocol is given by
\beq
\Gamma_{\text{C}}^{\textit{DF}} = \max \bigg\{ \Gamma_{\textit{SD}}^2+2\Gamma_{\textit{SD}}, \Gamma_{\text{eq}}^{\textit{DF}} \bigg\}.
\eeq
It can be shown that the pdf of $\Gamma_{\text{C}}^{\textit{DF}}$ under the condition $\Gamma_{\textit{SD}}^2 + 2\Gamma_{\textit{SD}} > \Gamma_{\text{eq}}^{\textit{DF}}$ is given by
\beq
\label{PDF_DF_direct}
f_{\Gamma_{\text{C}}^{\textit{DF}}} (\gamma \mid \Gamma_{\textit{SD}}^2 + 2\Gamma_{\textit{SD}} > \Gamma_{\text{eq}}^{\textit{DF}}) = \frac{ \overline{\gamma}_{\textit{SD}} \cdot f_{\Gamma_{\textit{SD}}} (\xi) \cdot F_{\Gamma^{\textit{DF}}_{\text{eq}}}(\gamma)  }{2(\xi +1) \cdot (\overline{\gamma}_{\textit{SD}} -  \mathfrak{D}(\alpha_{1},\alpha_{2}))},
\eeq
where $\xi = \sqrt{\gamma+1} -1$.
Substituting \eqref{PDF_DF_direct} into \eqref{genavgCapacity} and making some manipulations, we can obtain the average ergodic capacity of direct transmission as
\begin{align}
\label{avg_direct_capacity}
\nonumber
\overline{\mathbf{\text{C}}}_\text{d}^{\textit{DF}}  
&= \frac{1}{\ln 2} \cdot \frac{1}{\overline{\gamma}_{\textit{SD}} - \mathfrak{D}(\alpha_{1},\alpha_{2})}    \bigg\{ \overline{\gamma}_{\textit{SD}} \cdot e^{\frac{1}{\overline{\gamma}_{\textit{SD}}}} \operatorname{E}_1(\frac{1}{\overline{\gamma}_{\textit{SD}}}) \\
&\phantom{======} -  \int_{0}^{\infty} \hspace{-7pt}  \ln(1+t) e^{-\alpha_{1}t^2 - \alpha_2 t}  \,  \mathrm{d} t \bigg\}.
\end{align}
Following the same procedure, we can arrive at the pdf of $\Gamma_{\text{C}}^{\textit{DF}}$ under the condition $\Gamma_{\textit{SD}}^2 + 2\Gamma_{\textit{SD}} < \Gamma_{\text{eq}}^{\textit{DF}}$ as
\beq
\label{PDF_DF_relay}
f_{\Gamma_{\text{C}}^{\textit{DF}}} (\gamma \mid \Gamma_{\textit{SD}}^2 + 2\Gamma_{\textit{SD}} < \Gamma_{\text{eq}}^{\textit{DF}}) = \frac{ \overline{\gamma}_{\textit{SD}} \cdot f_{\Gamma^{\textit{DF}}_{\text{eq}}} (\gamma) \cdot F_{\Gamma_{\textit{SD}}}(\xi)  }{\mathfrak{D}(\alpha_{1},\alpha_{2})}.
\eeq
It follows that the average ergodic capacity of relay transmission is given by
\begin{align}
\label{avg_relay_capacity}
\nonumber
\overline{\mathbf{\text{C}}}_\text{r}^{\textit{DF}} &= \frac{1}{\ln 2 \cdot  \mathfrak{D}(\alpha_{1},\alpha_{2})} \bigg\{ \overline{\gamma}_{\textit{SD}} \cdot e^{\alpha_{1}} \operatorname{E}_1(\alpha_{1}) \\
&\phantom{====}- \int_{0}^{\infty} \ln(1+t) \cdot e^{-\alpha_{1}t - \frac{1}{\overline{\gamma}_{\textit{SD}}}(\sqrt{1+t}-1)} \mathrm{d} t \bigg\}.
\end{align}

\subsection{AF Relaying Protocol}
With AF relaying protocol, the pdf of the equivalent received SNR over relay link, $\Gamma_{\text{eq}}^{\textit{AF}}$, is given by \eqref{PDF_Gamma_eq_AF}.
Substituting \eqref{PDF_Gamma_eq_AF} into \eqref{genPrdirect}, we can obtain the probability that the source node performs direct transmission, denoted by $\mathcal{P}_{\text{d}}^{\textit{AF}}$, i.e.,
\beq
\label{genPrdirectAF}
\mathcal{P}_{\text{d}}^{\textit{AF}} = 1 - \frac{1}{\overline{\gamma}_{\textit{SD}}} \mathfrak{A} (\beta_1,\beta_2),
\eeq
where $\mathfrak{A} (\beta_1,\beta_2)$ is defined to be
\beq
\mathfrak{A} (\beta_1,\beta_2) \triangleq \int_{0}^{\infty} 2\beta_1 (t^2+2t) e^{-\beta_2 (t^2+2t)} \operatorname{K}_1(2\beta_1 (t^2+2t)) \, \mathrm{d} t,
\eeq
where $\beta_1 = \frac{1}{\sqrt{\overline{\gamma}_{\textit{SR}} \cdot \overline{\gamma}_{\textit{RD}}}}$, $\beta_2 = \frac{1}{\overline{\gamma}_{\textit{SD}}} + \frac{1}{\overline{\gamma}_{\textit{SR}}} + \frac{1}{\overline{\gamma}_{\textit{RD}}}$.

The equivalent SNR of the three-node cooperative transmission with AF protocol is given by
\beq
\label{Gamma_AF}
\Gamma_{\text{C}}^{\textit{AF}} = \max \bigg\{ \Gamma_{\textit{SD}}^2+2\Gamma_{\textit{SD}},  \Gamma_{\text{eq}}^{\textit{AF}}  \bigg\}.
\eeq
The pdf of $\Gamma_{\text{C}}^{\textit{AF}}$ under the condition $\Gamma_{\textit{SD}}^2 + 2\Gamma_{\textit{SD}} > \Gamma_{\text{eq}}^{\textit{AF}}$ can be obtained as
\beq
\label{PDF_AF_direct}
f_{\Gamma_{\text{C}}^{\textit{AF}}} (\gamma \mid \Gamma_{\textit{SD}}^2 + 2\Gamma_{\textit{SD}} > \Gamma_{\text{eq}}^{\textit{AF}}) = \frac{ \overline{\gamma}_{\textit{SD}} \cdot f_{\Gamma_{\textit{SD}}} (\xi) \cdot F_{\Gamma^{\textit{AF}}_{\text{eq}}}(\gamma)  }{2(\xi +1) \cdot (\overline{\gamma}_{\textit{SD}} -  \mathfrak{A}(\beta_{1},\beta_{2}))}.
\eeq
Correspondingly, we can calculate the average ergodic capacity of direct transmission $\overline{\mathbf{\text{C}}}_\text{d}^{\textit{AF}}$ by averaging the instantaneous capacity over the pdf of $\Gamma_{\text{C}}^{\textit{AF}}$ under the condition $\Gamma_{\textit{SD}}^2 + 2\Gamma_{\textit{SD}} > \Gamma_{\text{eq}}^{\textit{AF}}$ given in \eqref{PDF_AF_direct}, as
\beq
\overline{\mathbf{\text{C}}}_\text{d}^{\textit{AF}} = \int_{0}^{\infty} \frac12 \log_2 \big(1 + \gamma \big) \cdot \frac{ \overline{\gamma}_{\textit{SD}} \cdot f_{\Gamma_{\textit{SD}}} (\xi) \cdot F_{\Gamma^{\textit{AF}}_{\text{eq}}}(\gamma)  }{2(\xi +1) \cdot (\overline{\gamma}_{\textit{SD}} -  \mathfrak{A}(\beta_{1},\beta_{2}))} \, \mathrm{d} \gamma.
\eeq
Similarly, the pdf of $\Gamma_{\text{C}}^{\textit{AF}}$ under the condition $\Gamma_{\textit{SD}}^2 + 2\Gamma_{\textit{SD}} < \Gamma_{\text{eq}}^{\textit{AF}}$ is given by
\beq
\label{PDF_AF_relay}
f_{\Gamma_{\text{C}}^{\textit{AF}}} (\gamma \mid \Gamma_{\textit{SD}}^2 + 2\Gamma_{\textit{SD}} < \Gamma_{\text{eq}}^{\textit{AF}}) = \frac{ \overline{\gamma}_{\textit{SD}} \cdot f_{\Gamma^{\textit{AF}}_{\text{eq}}} (\gamma) \cdot F_{\Gamma_{\textit{SD}}}(\xi)  }{\mathfrak{A}(\beta_{1},\beta_{2})},
\eeq
which can be applied to the calculation of $\overline{\mathbf{\text{C}}}_\text{r}^{\textit{AF}}$.
The resulting expression is omitted for conciseness.

\subsection{Numerical Examples}

\begin{figure}[h]
\begin{center}
    \includegraphics[width = 3 in] {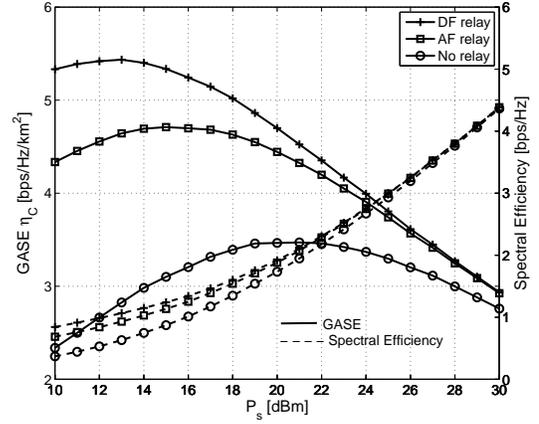}\\
\end{center}
\caption{The effect of the source node transmission power $P_{S}$ on GASE with DF and AF relaying protocol. ($P_{R} = 10$ dBm, $d_{\textit{SD}} = 1000$ m, $d_{\textit{SR}} = d_{\textit{RD}} = 500$ m, $\theta = 0$, $P_{\text{min}} = -80$ dBm, $N = -100$ dBm, $a = 4$.)}
\label{Power_DF}
\end{figure}

In Fig. \ref{Power_DF} we plot the GASE and spectral efficiency as function of the source node transmission power $P_{S}$ for DF and AF relaying protocol.
For comparison, we include the GASE curve of conventional point-to-point transmission without relays.
It shows that the cooperative transmission always enjoy better GASE performance than its conventional counterpart.
Meanwhile, the DF relaying protocol has slightly better overall performance than AF relaying protocol.
However, this performance gain shrinks as the transmission power $P_{S}$ increases.
Unlike spectral efficiency, whose performance curves are monotonically increasing function with respect to $P_{S}$, the GASE curves show a peak as transmission powers increase.
This observation indicates that increasing the transmission power can lead to a higher spectral efficiency but can not necessarily increase GASE.
Therefore, GASE provide a new perspective on transmission power selection for wireless transmitters.
Another interesting observation is that the optimal transmission power, in terms of maximizing the GASE, for point-to-point transmission is much larger than that for cooperative relay transmission. In summary, cooperative relay transmission can enjoy much higher energy efficiency than point-to-point transmission when the relay is properly located.

\section{GASE of Underlay Cognitive Radio Transmission}

\begin{figure}[h]
\begin{center}
    \includegraphics[width = 2 in] {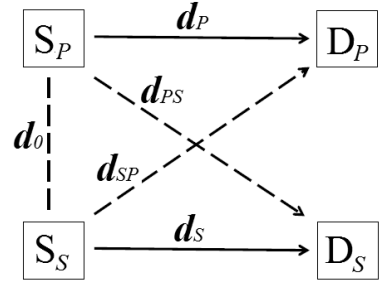}\\
\end{center}
\caption{System model of underlay cognitive radio transmission. $\text{S}_P$, $\text{S}_S$, $\text{D}_P$, $\text{D}_S$ represent a primary transmitter, a secondary transmitter, a primary receiver and a secondary receiver, respectively.}
\label{Cognitive}
\end{figure}

So far, we have studied the scenarios where only a single transmitter is operating at any time.
In this section, we generalize the analysis to consider the case where parallel transmission occurs.
Specifically, we extend the GASE analysis to underlay cognitive radio transmission and examine the effect of interference on overall spectral utilization efficiency while taking into account the larger spatial footprint of parallel transmission. We consider the transmission scenario as illustrated in Fig. \ref{Cognitive}.
The primary user $\text{S}_P$ transmits to the primary receiver $\text{D}_P$ with transmission power $P_1$.
Meanwhile, the secondary user $\text{S}_S$ opportunistically communicates with the secondary receiver $\text{D}_S$ using the same frequency bandwidth with transmission power $P_2$. Both transmitters use omni-directional antennas.
As such, the primary user (secondary user) will generate interference on the secondary receiver (primary receiver).
The distance of the transmission link between $\text{S}_P$ ($\text{S}_S$) and $\text{D}_P$ ($\text{D}_S$) is denoted as $d_{\textit{P}}$ ($d_{\textit{S}}$); whereas the distance of the interference link between $\text{S}_P$ ($\text{S}_S$) and $\text{D}_S$ ($\text{D}_P$) is denoted as $d_{\textit{PS}}$ ($d_{\textit{SP}}$).
The distance between $P_T$ and $S_T$ is denoted as $d_0$.
In the underlay paradigm, the secondary user is allowed to utilize the primary user's spectrum as long as the interference it generates on the primary receiver is below a pre-determined threshold $I_{th}$.
Otherwise, the secondary user should remain silent. We assume that, by exploring the channel reciprocity, the secondary transmitter can predict the amount of interference that its transmission will generate to the primary receiver.

We first focus on the scenario where parallel secondary transmission occurs, i.e., the received interference power from secondary transmitter at the primary receiver is less than  $I_{th}$.
Based on the path loss and fading model adopted in this work, this interference power is given by $P_2 \cdot Z / d_{\textit{SP}}^{a}$.
As such, the probability that the parallel transmission occurs is then given by $\mathcal{P} = \mathbb{P} \left\{ P_2 \cdot Z / d_{\textit{SP}}^{a} < I_{th} \right\}$.
Under Rayleigh fading environment, this probability specializes to
\beq
\label{Probparellel}
\mathcal{P} = 1 - \exp \left(- \frac{I_{th} \cdot d_{\textit{SP}}^{a}}{P_2}\right).
\eeq

\subsection{Ergodic capacity analysis}
In the underlay cognitive transmission scenario, the total instantaneous capacity of both primary and secondary transmissions is given by
$
\text{C}_{\textit{CR}} = \log_2\left(1 + \Gamma_p \right) + \log_2\left(1 + \Gamma_s \right),
$
where $\Gamma_p$ and $\Gamma_s$ denote the received signal-to-interference-plus-noise ratio (SINR) at primary receiver $\text{D}_P$ and secondary receiver $\text{D}_S$, respectively. Based on the adopted path loss and fading models, $\Gamma_p$ and $\Gamma_s$ can be shown to be given by
\begin{equation}
\Gamma_p= \frac{P_1 \cdot Z_{\textit{P}} / d_{\textit{P}}^a }{P_2 \cdot Z_{\textit{SP}} / d_{\textit{SP}}^a + N},\ \ \Gamma_s= \frac{P_2 \cdot Z_{\textit{S}} / d_{\textit{S}}^a }{P_1 \cdot Z_{\textit{PS}} / d_{\textit{PS}}^a + N},
\end{equation}
respectively.
It follows that the ergodic capacity of the parallel transmission channel can be calculated as
\beq
\label{CX}
\overline{\text{C}}_{\textit{CR}} = \underbrace{\int_{0}^{\infty} \hspace{-5pt} \log_2 (1+\gamma) \cdot \mathrm{d} F_{\Gamma_p} (\gamma) }_{\overline{\text{C}}_{\textit{CR}}^p} + \underbrace{\int_{0}^{\infty} \hspace{-5pt} \log_2 (1+\gamma) \cdot \mathrm{d} F_{\Gamma_s} (\gamma) }_{\overline{\text{C}}_{\textit{CR}}^s},
\eeq
where $F_{\Gamma} (\cdot)$ denotes the cdf of SINR $\Gamma$.

Under Rayleigh fading environment, the cdf of $\Gamma_p$, $F_{\Gamma_p} (\cdot)$, can be derived, while considering the interference constraint on $\text{D}_P$, as
\begin{align}
\label{CDFGammaprimaryuser}
\nonumber
\hspace*{-3pt} F_{\Gamma_p}  (& \gamma )  =  1 - \exp \left( -\frac{d_{\textit{SP}}^a \cdot I_{th}}{P_2} \right) - \frac{\rho_p}{\gamma + \rho_p} \exp \left( -\frac{d_{\textit{P}}^a \cdot N}{P_1} \gamma \right) \\
& \times \left\{ 1 - \exp \left( -\frac{d_{\textit{SP}}^a \cdot I_{th}}{P_2} \right) \exp \left( -\frac{d_{\textit{P}}^a \cdot I_{th}}{P_1} \gamma \right)  \right\},
\end{align}
where $\rho_p = \frac{P_1}{P_2} \left( \frac{d_{\textit{SP}}}{d_{\textit{P}}} \right)^a$.
It follows that the ergodic capacity of the primary user in the parallel channel over Rayleigh fading can be calculated, after substituting \eqref{CDFGammaprimaryuser} into the first part of \eqref{CX} and applying integration by part, as
\begin{equation}
\label{Cprimaryuser}
\overline{\text{C}}_{\textit{CR}}^{p} =
\begin{cases}
\frac{1}{\ln2} \frac{\rho_p}{1 - \rho_p}  \bigg\{ \mathfrak{F} \left( \frac{d_{\textit{P}}^{a} N}{P_1} \rho_p \right) - \mathfrak{F} \left( \frac{d_{\textit{P}}^{a} N}{P_1} \right) - \exp\left( - \frac{d_{\textit{SP}}^{a} I_{th}}{P_2} \rho_p \right) \\
\qquad \times \left\{ \mathfrak{F} \left( \frac{d_{\textit{P}}^{a} (N+I_{th})}{P_1} \rho_p \right) - \mathfrak{F} \left( \frac{d_{\textit{P}}^{a} N}{P_1} \right)\right\}, \qquad \rho_p \neq 1, \\
\frac{1}{\ln2} \bigg\{ 1 - \frac{d_{\textit{P}}^{a} N}{P_1} \mathfrak{F} \left( \frac{d_{\textit{P}}^{a} N}{P_1}\right) - \exp \left( - \frac{d_{\textit{SP}}^{a} I_{th}}{P_2} \right) \\
\qquad \times \left\{ 1 - \frac{d_{\textit{P}}^{a} (N+I_{th})}{P_1} \mathfrak{F} \left( \frac{d_{\textit{P}}^{a} (N+I_{th})}{P_1}\right) \right\} \bigg\},  \quad \, \, \rho_p = 1,
\end{cases}
\end{equation}
where $\mathfrak{F} (x) \triangleq \exp(x)\cdot \operatorname{E_1} (x)$.

The cdf of $\Gamma_s$ can be similarly obtained, but without the interference power constraint, as
\beq
\label{CDFGammaseconduser}
F_{\Gamma_{s}}(\gamma) = 1 - \frac{\rho_s}{\gamma+\rho_s} \exp \left( -\frac{d_{\textit{S}}^{a} N}{P_2} \gamma\right),
\eeq
where $\rho_s = \frac{P_2}{P_1} \left( \frac{d_{\textit{PS}}}{d_{\textit{S}}} \right)^a$.
Finally, the ergodic capacity of the secondary user with the presence of primary user interference is given by
\beq
\label{Cseconduser}
\overline{\text{C}}_{\textit{CR}}^{s} =
\begin{cases}
\frac{1}{\ln2} \frac{\rho_s}{1 - \rho_s} \bigg\{ \mathfrak{F} \left( \frac{d_{\textit{S}}^{a} N}{P_2} \rho_s \right) - \mathfrak{F} \left( \frac{d_{\textit{S}}^{a} N}{P_2} \right) \bigg\}, & \rho_s \neq 1,\\
\frac{1}{\ln2} \bigg\{ 1 - \frac{d_{\textit{S}}^{a} N}{P_2} \mathfrak{F} \left( \frac{d_{\textit{S}}^{a} N}{P_2}\right)\bigg\}, & \rho_s = 1.
\end{cases}
\eeq

\subsection{Affected area analysis}

Based on the definition of affected area adopted in this work, a particular area is affected if the total received signal power from both transmitters is greater than $P_{\text{min}}$.
Specifically, the probability that an incremental area of distance $r_p$ to the primary transmitter $\text{S}_P$ and distance $r_s$ to secondary transmitter $\text{S}_S$ is affected can be calculated as the probability that the total received signal power $P_r(r_p) + P_r(r_s)$ is greater than $P_{\text{min}}$, i.e., $\mathbb{P} \bigg\{ P_r(r_p) +P_r(r_s) \ge P_{\text{min}} \bigg\}$.
It follows that the affected area of parallel transmission can be calculated as
\begin{equation}
\label{AaffgenXchannel}
\text{A}_{\textit{CR}}^{pt} = \int_0^{2\pi}\int_{0}^{\infty} \mathbb{P} \bigg\{ P_r(r_p) +P_r(r_s) \ge P_{\text{min}} \bigg\} r_p\mathrm{d} r_p \mathrm{d} \theta,
\end{equation}
where $r_s = \sqrt{r_p^2 + d_0^2 - 2 r_p d_0\cos{\theta}}$.
Note that we use the location of $\text{S}_P$ as the origin in the about integration.

Under Rayleigh fading environment, the pdf of the total received signal power at a location of distance $r_p$ to transmitter $\text{S}_P$ and distance $r_s$ to transmitter $\text{S}_S$, $X = P_r(r_p) + P_r(r_s)$, can be obtained as
\beq
f_X(x)=
\begin{cases}
\frac{1}{\lambda_p - \lambda_s} \bigg( e^{-\lambda_p x} - e^{-\lambda_s x}  \bigg), & \lambda_p \neq \lambda_s,\\
\frac{x}{\lambda_p^2} e^{-\lambda_p x}, & \lambda_p = \lambda_s,
\end{cases}
\eeq
where $\lambda_i = P_i / r_i^a$, $i \in \{p,s\}$.
Accordingly, the probability that this location is affected is determined as
\begin{align}
\label{Xchannelaff}
\nonumber
& \mathbb{P} \bigg\{ P_r(r_p) + P_r(r_s) \ge P_{\text{min}} \bigg\} \\
= &
\begin{cases}
\frac{1}{1-\lambda_p/\lambda_s} e^{-P_{\text{min}}/\lambda_p} + \frac{1}{1-\lambda_s/\lambda_p} e^{-P_{\text{min}}/\lambda_s}, &  \lambda_p \neq \lambda_s \\
( 1 + P_{\text{min}} / \lambda_p ) e^{- P_{\text{min}} / \lambda_p} , & \lambda_p = \lambda_s.
\end{cases}
\end{align}
Substituting \eqref{Xchannelaff} into \eqref{AaffgenXchannel}, we can numerically calculate the affected area of underlay cognitive radio transmission with parallel transmission, $\text{A}_{\textit{CR}}^{pt}$.
The GASE of parallel transmission over Rayleigh fading channel can be calculated as
\beq
\eta_{\textit{CR}}^{pt} = \frac{\overline{\text{C}}_{\textit{CR}}^{p} + \overline{\text{C}}_{\textit{CR}}^{s}}{\text{A}_{\textit{CR}}^{pt}}.
\eeq

When the parallel secondary transmission is prohibited, i.e., the interference power constraint $P_2 \cdot Z / d_{\textit{SP}}^{a} < I_{th}$ on $\text{D}_P$ is not satisfied, the secondary user is not allowed to utilize the primary user's spectrum. Therefore, the parallel transmission channel simplifies to the point-to-point link, whose GASE $\eta_{\textit{CR}}^{st}$ is given by \eqref{GASEforsingleuserray} with the transmission power $P_t$ and distance $d$ substituted by $P_1$ and $d_{\textit{P}}$, respectively.
Finally, GASE for underlay cognitive radio transmission can be written as
\beq
\label{GASEcognitive}
\eta_{\textit{CR}}^{} = \mathcal{P} \cdot \eta_{\textit{CR}}^{pt} + (1 - \mathcal{P}) \cdot \eta_{\textit{CR}}^{st}.
\eeq
Note that the GASE expression derived above for underlay cognitive radio transmission will reduce to that for the $X$ channels, resulted from insufficient spatial separation between receivers \cite{jafar08, meddahali08, Jafarkhani12}, when the interference threshold approaches infinity. In particular, when $I_{th} \rightarrow \infty$, $\mathcal{P}$ approaches to $1$, which means two transmitters, $\text{S}_P$ and $\text{S}_S$, always transmit simultaneously over the same frequency band.
The GASE of $X$ channels is then given by \cite[eq. 17]{Zhang2012}
\beq
\label{GASEX}
\eta_{\textit{X}} = \frac{\overline{\text{C}}_{\textit{CR}}^{p'} + \overline{\text{C}}_{\textit{CR}}^{s}}{\text{A}_{\textit{CR}}^{pt}},
\eeq
where
\beqn
\label{Cprimaryuserinfty}
&&\overline{\text{C}}_{\textit{CR}}^{p'} = \lim_{I_{th} \rightarrow \infty} \overline{\text{C}}_{\textit{CR}}^{p} \nonumber \\&&=
\begin{cases}
\frac{1}{\ln2} \frac{\rho_p}{1 - \rho_p} \bigg\{ \mathfrak{F} \left( \frac{d_{\textit{P}}^{a} N}{P_1} \rho_p \right) - \mathfrak{F} \left( \frac{d_{\textit{P}}^{a} N}{P_1} \right) \bigg\}, & \rho_p \neq 1,\\
\frac{1}{\ln2} \bigg\{ 1 - \frac{d_{\textit{P}}^{a} N}{P_1} \mathfrak{F} \left( \frac{d_{\textit{P}}^{a} N}{P_1}\right) \bigg\},  & \rho_p = 1.
\end{cases}
\eeqn
and $\overline{\text{C}}_{\textit{CR}}^{s}$ and $\text{A}_{\textit{CR}}^{pt}$ are given in \eqref{Cseconduser} and \eqref{AaffgenXchannel}, respectively.

\subsection{Numerical examples}
\begin{figure}[h]
\begin{center}
    \includegraphics[width = 3 in] {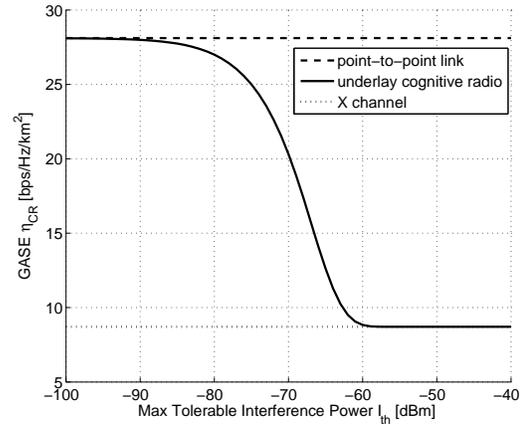}\\
\end{center}
\caption{The effect of the max tolerable interference power $I_{th}$ on GASE. $P_1 = P_2 = 20$ dBm, $N = -100$ dBm, $P_{\text{min}} = -100$ dBm, $a = 4$, $d_0 = d_{\textit{P}} = d_{\textit{S}} = 100$ m, $\kappa = 1.5$.}
\label{interfmax}
\end{figure}

In Fig. \ref{interfmax}, we investigate the effect of the maximum tolerable interference power $I_{th}$ on GASE performance of underlay cognitive radio systems. To simplify the notation and discussion, we denote the ratio of the interfering link distance to transmission link distance for primary receiver by $\kappa_p= d_{\textit{SP}} / d_{\textit{P}}$ and that for the secondary receiver by $\kappa_s= d_{\textit{PS}} / d_{\textit{S}}$.
Note that the distance of the interfering link between $\text{S}_S$ and $\text{D}_P$, $d_{\textit{SP}}$, should satisfy
$|d_0 - d_{\textit{P}}| < d_{\textit{SP}} < |d_0 + d_{\textit{P}}|$.
Therefore, $\kappa_p = d_{\textit{SP}} / d_{\textit{P}}$ should be bounded as
$|d_0 / d_{\textit{P}} - 1| < \kappa_p < |d_0 / d_{\textit{P}} + 1|$.
Similar bound applies to $\kappa_s$.
In Fig. \ref{interfmax}, we plot GASE of underlay cognitive radio transmission as function of $I_{th}$ while fixing $\kappa_p = \kappa_s = \kappa = 1.5$. We can see that as $I_{th}$ decreasing, the GASE performance of underlay cognitive radio transmission converges to that of the point-to-point transmission case.
This behavior can be explained that when $I_{th} \rightarrow 0$, the probability of parallel transmission $\mathcal{P}$ given in \eqref{Probparellel} approaches to $0$.
On the other hand, when $I_{th} \rightarrow \infty$, the GASE performance of underlay cognitive radio transmission converges to that of $X$ channels, as expected by intuition. We also notice that the GASE performance of underlay cognitive radio transmission is always worse than that of point-to-point transmission case but better than $X$ channel transmission for the chosen system parameters. In the next numerical example, we explore under what scenario underlay cognitive transmission will lead to better GASE performance.

\begin{figure}[h]
\centering

\subfigure[Spectral Efficiency]
{
    \includegraphics[width = 3 in] {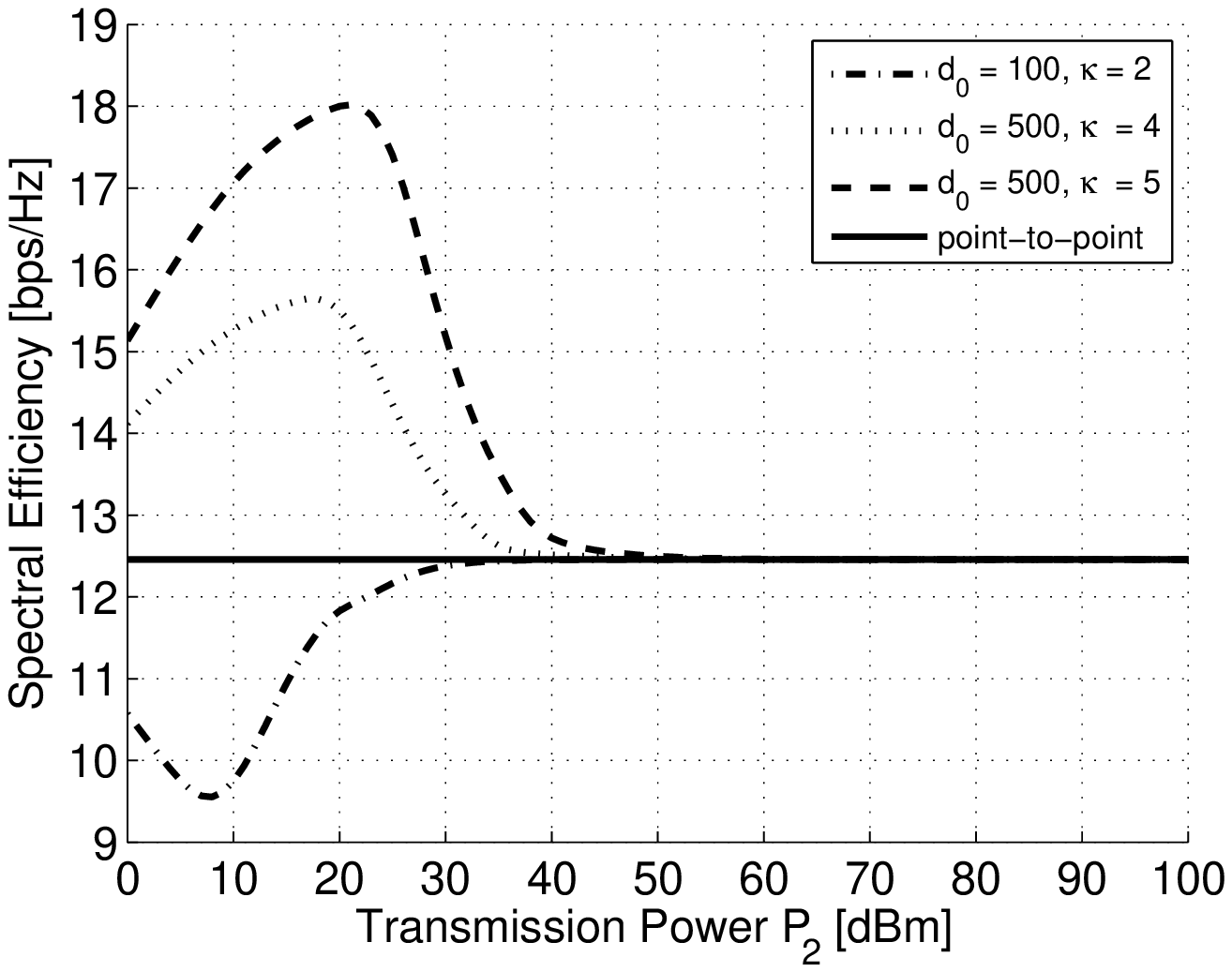}
    \label{fixedpt1SE}
}

\subfigure[GASE]
{
    \includegraphics[width = 3 in] {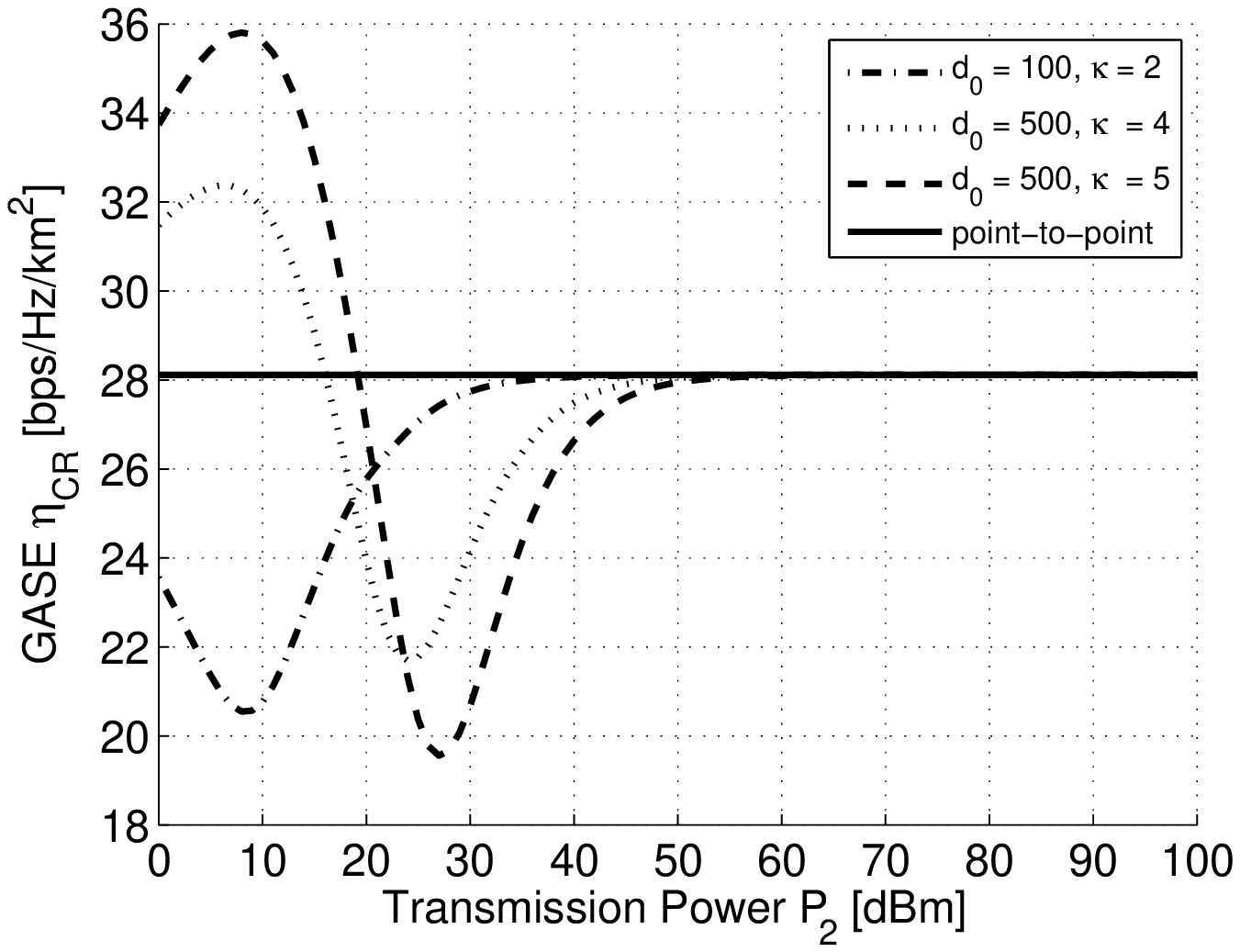}
    \label{fixedpt1}
}
\caption{The effect of the transmission power of secondary user $P_2$ on the system overall spectral efficiency and GASE. $P_1 = 20$ dBm, $N = -100$ dBm, $P_{\text{min}} = -100$ dBm, $I_{th} = -80$ dBm, $a = 4$, $d_{\textit{P}} = d_{\textit{S}} = 100$ m.}
\label{SEGASE}
\end{figure}

In Fig. \ref{SEGASE}, we compare GASE metric with conventional spectral efficiency of underlay cognitive radio systems.
Specifically, we plot overall spectral efficiency in Fig. \ref{fixedpt1SE} and GASE  in Fig. \ref{fixedpt1} as function of the transmission power of secondary user $P_2$ for different system parameters.
As we can see, when the interfering transmitter is close to the target receiver, i.e., $\kappa$ is small, introducing underlay cognitive transmission always deteriorates the overall system spectral efficiency as well as the GASE performance.
Basically, capacity gain incurred by the spectrum sharing through parallel transmission cannot compensate the capacity loss caused by the mutual interference, even with underlaying interference threshold requirement.
The interference requirement at the primary user may protect the primary user transmission but the secondary transmission will suffer severe interference from primary user transmission.
This observation also justifies the practical understanding that the radio spectrum should not be simultaneously used by other transmissions very close to the transmitter and/or the receiver.
On the other hand, when $\kappa$ is large, the underlay cognitive radio transmission results in different behaviors in terms of spectral efficiency and GASE.
First, the underlay cognitive radio transmission always benefits from the secondary user transmission in terms of spectral efficiency when the transmission power is not very large, as shown in Fig. \ref{fixedpt1SE}.
However, with respect to the GASE performance metric, the underlay cognitive radio transmission may benefit from the secondary user only if its transmission power is carefully selected.
Otherwise, the overall GASE performance may be greatly deteriorated by the secondary cognitive transmission, as shown in Fig. \ref{fixedpt1}.
Therefore, subject to GASE performance metric, the power allocation for spectrum sharing transmission should be carefully designed.
Finally, both of the spectral efficiency and GASE curve show the same asymptotic approach to point-to-point link as increasing $P_2$.
This is because when $P_2 \rightarrow \infty$, the probability of parallel transmission $\mathcal{P}$ given in \eqref{Probparellel} approaches to $0$.
Under this circumstance, the performance of the cognitive radio transmission approaches to that of the point-to-point link, as shown in the large $P_2$ region in Fig. \ref{SEGASE}.

\section{Conclusion}
In this paper, we generalized the conventional ASE performance metric to study the performance of arbitrary wireless transmissions while considering the spatial effect of wireless transmissions.
We carried out a comprehensive study on the resulting GASE performance metric by considering point-to-point transmission, dual-hop relay transmission, cooperative relay transmission as well as cognitive radio transmission.
Through analytical results and selected numerical examples, we showed that our research provided a new perspective on the design, evaluation and optimization of arbitrary wireless transmissions, especially with respect to the transmission power selection.
Meanwhile, we show that relay transmission is power efficient and can greatly improve the greenness of wireless transmissions.
Finally, the study on underlay cognitive radio transmission implied that, if the power of secondary transmitter is not properly chosen, the secondary user may degrade the GASE performance of overall system, even worse than that of the point-to-point primary transmission only case.
While the analysis focuses on single antenna per node scenario, the generalization of the analysis to multiple antenna cases is straightforward.
The GASE metric can also apply to the spectral utilization efficiency of wireless ad hoc network and femtocell enhanced cellular systems after proper adaptation.

\end{document}